# PREDICTING GROWTH AND FINDING BIOMASS PRODUCTION USING THE GENERAL GROWTH MECHANISM

YURI K. SHESTOPALOFF

SegmentSoft Research Lab, 61 Danby Ave, Toronto, Ontario, M3H 2J4 Canada
shes169@yahoo.ca

**Abstract**. First, we briefly describe the general growth mechanism, which governs the growth of living organisms, and its mathematical representation, the growth equation. Using the growth equation, we compute growth curve for *S. cerevisiae* and show that it corresponds to available experimental data. Then, we propose a new method for finding amount of synthesized biomass without complicated stoichiometric computations and apply this method to evaluation of biomass production by *S. cerevisiae*. We found that obtained results are very close to values obtained by methods of metabolic flux analysis. Since methods of metabolic flux analysis require finding produced biomass, which is one of the most important parameters affecting stoichiometric models, a priori knowledge of produced biomass can significantly improve methods of metabolic flux analysis in many aspects, which we also discuss. Besides, based on the general growth mechanism, we considered evolutionary development of *S. cerevisiae* and found that it is a more ancient organism than *S. pombe* and is apparently its direct predecessor.

## 1. Introduction

In this paper, we apply the general law of growth and replication[1] for growth prediction and computing the amount of synthesized biomass.

The core essence of the general growth law is that it regulates the distribution of incoming nutritional resources between the maintenance needs of organisms and biomass synthesis in a certain predefined way that depends on geometrical characteristics and biochemical machinery of organisms. In a simple form, it can be formulated as follows: "*The growth cycle of a living organism and appropriate progression in composition of biochemical reactions are defined by the distribution of nutrients between maintenance needs and biomass production in such a way that the fraction of nutrients directed to biomass production at any given moment is equal to the growth ratio (which is a monotonically decreasing function), so that the growth rate is proportional to influx of nutrients and the growth ratio.*"

It can be viewed as a common sense consideration that when some organism grows, it requires more resources for maintenance, just because of the size increase. Since the amount of nutrients is restricted, then during growth less and less nutrients can be allocated for organism's increase (biomass production). It turns out that this distribution of resources is a well defined process. The general growth law and its mathematical representation, the growth equation, describe this process in mathematical terms.

The parameter "growth ratio" is not a magic number but a mathematical representation of a certain Nature's mechanism that optimizes the use of supplied nutrients between the need to *maintain* certain biomass (associated with volume) and to *synthesize* new biomass. The growth ratio depends on the geometrical characteristics and, indirectly, on biochemical properties of organisms.

When an organism grows and its mass increases, the relative fraction of incoming nutrients that is diverted to maintenance of existing biomass increases, and accordingly the relative fraction of nutrients that is used by the organism for biomass synthesis decreases. The biochemical machinery changes accordingly, in order to correspond to a modified distribution of nutrients between maintenance needs and a reduced amount of synthesized biomass.

Since growth is a complex multifactor phenomenon, the workings of the general growth law present themselves through different, more specific, mechanisms, such as biochemical reactions and geometrical and physical factors. These "second line" mechanisms interrelate and work in close cooperation under the guidance of the general growth law.

In this paper, using the general growth mechanism, and its mathematical representation, the growth equation, we propose a simple independent, self sufficient method for evaluation of biomass production using example of *S. cerevisiae*. Then, we compare these results to similar data obtained by far more complicated methods of metabolic flux analysis. The new method can be beneficially used for many purposes, including significant improvement of methods of metabolic flux analysis, both from computational and application perspectives, since amount of synthesized biomass is a critical parameter for these methods.

**2. The growth equation**

Mathematically, the general growth law is represented by the growth equation. In simple growth scenarios, like the growth of a unicellular microorganism that receives nutrients through its membrane, the growth equation has the following form.

$$p_c(X)dV(X,t) = k(X,t) \times S(X) \times \left(\frac{R_S}{R_V} - 1\right)dt \qquad (1)$$

Here, $X$ is a vector that represents coordinates of some point in a cell; $p_c$ is the density of the cell's substance measured in $kg/m^3$ that can generally depend on the coordinate vector $X$; $t$ is time; $k$ is the specific influx; $p_c(X)dV(X,t)$ is the change in the cell's mass; $S(X)$ is the cell's surface area; $R_S$ and $R_V$ denote the relative surface and volume accordingly, which can also depend on the coordinate vector $X$ and time $t$. In general, the density of the cell $p_c$ can also depend on the coordinate vector $X$.

The growth equation has the following interpretation. The left part represents the mass increment. The right part represents the total influx of nutrients through the surface (the term $k(X,t) \times S(X)$), multiplied by the value of the *growth ratio* $G_R = (R_S / R_V - 1)$. It defines the *fraction* of the total influx that is used for biomass synthesis. For multicellular organisms, one should use the total influx; for instance, when nutrients come through a fruit stem or blood vessels. The growth ratio depends on the geometrical characteristics of an organism and, indirectly, on the organism's biochemical machinery through the *maximum possible size* that can be achieved for a particular organism.

As an example, let us consider a cell that has a disk like shape. We can find the relative surface, the relative volume and the growth ratio for the disk as follows.

$$R_S = \frac{S(V)}{S(V_0)} = \frac{r(r+H)}{R(R+H)} \qquad (2)$$

$$R_V = \frac{V}{V_0} = \frac{r^2}{R^2} \qquad (3)$$

$$G_R = \frac{R_S}{R_V} - 1 = \frac{R(r+H)}{r(R+H)} - 1 \qquad (4)$$

where $V_0$ is maximum volume; $S_0$ is surface area corresponding to $V_0$; $r$ is the current radius of the disk, $R$ is the radius of the disk corresponding to the maximum possible volume; $H$ is the disk's height (assumed to be constant).

**3. Influx of nutrients**

The authors in Ref. 2 discovered a *doubling* in the rate of rRNA synthesis and poly(A)-containing RNA in *S. cerevisiae* during S phase, and the preservation of this high rate through the growth cycle. The same results were reported for *S. pombe*[2, 3]. Since these results were obtained for *synchronically* growing microorganisms, they are applicable to an *individual* cell. At the same time, the rate of protein synthesis remains largely the same through the whole growth cycle. So, for *S. cerevisiae*, we can assume that the rate of RNA synthesis is double that of protein.

Rates of synthesis are directly related to the amount of nutrients required for the synthesis of particular components[1]; in other words, they are linked to nutrients' influx. For proteins, as a first approximation, we may use the law of conservation of mass and assume that the total influx that is used for protein synthesis (and for other substances synthesized at about the same rate as protein) is proportional to the cell mass. This approach is used in metabolic flux analysis[4, 5] and is consistent with the law of preservation of mass. On the other hand, the cell does many other things besides the synthesis of protein, such as supporting transportation and signaling networks (infrastructure costs), proofreading of DNA and protein, proton leakage across membrane, etc.[6]. All these numerous activities require energy and consequently nutrients.

Protein constitutes a relatively stable and also the largest part, about 55%, of the total cell mass. Since the rate of RNA synthesis is double that of protein, the influx of nutrients for synthesis of ribosomes is proportional to the square of mass. The efficiency of biochemical machinery is about the same for synthesis of proteins and RNA, because both processes use transcription and translation mechanisms of the same nature. So, we can assume that the influx of nutrients that is required for synthesis is proportional to relative contents of protein and RNA. The relative content of different components may remain the same through the whole growth cycle, although rates of synthesis may differ. This is the case for protein and RNA content in many organisms, in which the component that is synthesized with a higher rate decays faster. Using these considerations, we can define the specific influx required for protein and RNA synthesis.

Ref. 1 presents other proofs of existence of two typical biochemical arrangements. One is when the rates or protein and RNA synthesis are about the same. For instance, this is the case for *amoeba*. If we assume that the cell density is the same, we can substitute mass for volume. Then, the influx of nutrients can be defined as follows[1].

$$K_{\min} = \left(C_p v + C_r v\right) \quad (5)$$

Here, $C_r$ and $C_p$ are the weighting coefficients (units of measure $[kg/\sec]$) corresponding to fractions of influx that are used to synthesize protein and RNA. Note that we use the *dimensionless* volume *v* in (5), which is the ratio of the current volume to the *initial* volume, so that *v* is a dimensionless value greater than one.

The other typical biochemical machinery[1], which supports accelerated growth, is when the rate of RNA synthesis is about double the rate of protein synthesis, which is the case of *S. cerevisiae*, *S. pombe* and *E. coli*. For such organisms, we have

$$K_{\min}(v) = \left(C_p v + C_r v^2\right) \quad (6)$$

It is quite reasonable to assume that transportation and signaling costs are proportional to the distance the nutrients and the waste (which is associated with nutrients) have to be transported and the signals transmitted. Then, we can find the total influx that includes infrastructure costs as follows.

$$dK(L) = K_{\min}(L)dL \quad (7)$$

where *L* is a relative increase of network length.
This formula is a mathematical representation of our assumption that the amount of additional nutrients that are required to transport one unit of influx into the destination point of synthesis is proportional to the traveled distance. Substituting (6) into (7) and solving it, we find.

$$K(L) = \left(C_p L^2 + C_r L^3\right) \quad (8)$$

Similarly, we can consider two- and three-dimensional growth. For instance, for the disk, whose height remains constant during growth, and the rates of protein and RNA synthesis are the same, we find:

$$K = \frac{C}{H}\left(\frac{v^{3/2}}{\sqrt{H}}\right) \quad (9)$$

where $v = V/V_b$, and $V_b$ is the beginning disk volume; $C$ is a constant coefficient. Note that without infrastructure costs, influx is proportional to increase of relative volume.

Accounting for transportation and signaling costs, for a sphere we have $K = Cv^{4/3}$, when the rates of protein and ribosome synthesis are the same. If the rate of RNA synthesis is double that of protein, such as in the case of *S. cerevisiae*, then

$$K(V) = \left(C_p(V/V_b)^{4/3} + C_r(V/V_b)^{7/3}\right) \quad (10)$$

Generalizing these considerations, we can write the growth equation that takes into account the infrastructure "toll" for all cell components as follows.

$$p_c(X)dV(X) = k_{\min}(X) \times (r(X)/r_0(X))S(X) \times \left(\frac{R_S}{R_V} - 1\right)dt \quad (11)$$

where the new variable $r$ is the distance that the synthesized and raw substances have to be transported; $r_0$ is the initial transport distance in the same direction; $k_{\min}$ is the minimum required nutrients' influx without "infrastructure costs". Since much of the activity in a cell is directed from the periphery to the center and vice versa, in many instances it will be reasonable to assume that $r$ in (11) is the distance from the center to the elementary volume $dV(X)$.

**4. Computing S. cerevisiae's growth curve**

The growth equation (1) and (11) define the *possible* growth curve. It was discovered in Refs. 1, 7 that evolutionarily Nature developed at least two typical growth scenarios. One is when organisms use the whole growth cycle predefined by the growth equation, like *amoeba*. In the second growth scenario, fast growing organisms such as *S. pombe* use only the fastest part of the possible growth curve and then, through the use of more sophisticated biochemical machinery[8], switch to division phase at inflection point of the growth curve[1]. In this regard, *S. cerevisiae* exercises the first type of growth scenario, like *amoeba*. On the other hand, unlike *amoeba*, its biochemical machinery is similar to *S. pombe* in that the rate of RNA synthesis is about double the rate of protein synthesis.

If *S. cerevisiae* has a spherical shape, its influx of nutrients would be defined by equation (10). However, it has an ellipsoid shape, and so we should use the growth equation (11). Also, we need to know maximum possible size of *S. cerevisiae*. The growth of *S. cerevisiae* has some specifics, summarized Ref. 9 as follows: "*The growth in dry mass was linear for each budding cycle. The combined system of mother cell plus bud grows in dry mass at a constant rate from 20 minutes after the appearance of the bud until 20 minutes after the appearance of the next pair of buds. It is concluded that dry mass growth is linear for most of each cell generation. ... The curve for volume growth is approximately sigmoid.*"

In Ref. 10, the authors studied different aspects of *S. cerevisiae* growth, mostly the growth of populations. They discovered that these organisms grow in aggregations, which certainly influences growth processes. For populations, the growth curve also resembles a sigmoid shape, with a clear horizontal asymptote. Of course, the fact that the population's growth has S-curve type of growth is not a 100% guarantee that an individual cell has also S-curve type of growth. However, we found that there is some correlation between types of growth for a population and individual organisms and that the populations of organisms whose *individual* growth curves have S-curve and J-curve shapes, accordingly tend to have S- and J-shaped *population* growth curves.

Combining the results presented in the cited studies, we can assume that a single *S. cerevisiae* cell has a sigmoid growth curve. We model the geometrical form of *S. cerevisiae* by an ellipsoid of revolution that grows in all dimensions, but the rate of growth in 'width" and "length" can be different. Such a geometrical form is completely characterized by initial radius, initial ratio of axes, ending radius, and ending ratio of axes. We used photos of *S. cerevisiae* available on the internet and in Ref. 11, in order to calculate the range of initial and ending ratios of axes. An example of geometrical characteristics that we used in computations is presented in table 1.

Table 1. Example of geometrical characteristics for modeling *S. cerevisiae*.

| Geometrical form | Initial radius | Radius Increase (times) | Initial axes ratio | Ending axes ratio |
|---|---|---|---|---|
| 1. close to sphere | 1.0 | 1.267 | 1.01 | 1.02 |
| 2. ellipsoid | 1.0 | 1.223 | 1.23 | 1.35 |
| 3. elongated ellipsoid | 1.0 | 1.02 | 1.8 | 3.46 |

The ellipsoid's cross section along the larger axis is shown in Fig. 1.

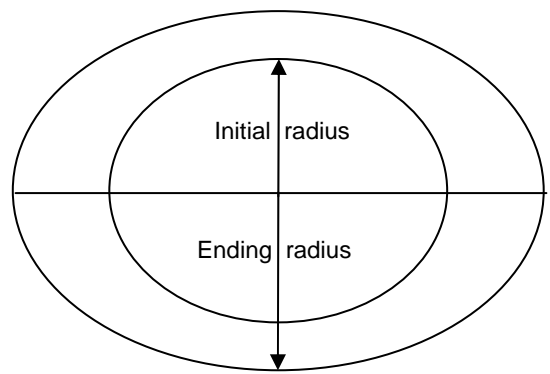

Fig. 1. Geometrical form of growing *S. cerevisiae*.

There are specific issues with regard to growth of *S. cerevisiae* that have to be taken into account when we interpret experimental data. First, the author of Ref. 9 measured only mass. Volume was not measured but *calculated* based on the assumption that the cell is a prolate ellipsoid of revolution, so that the volume was computed using formulas for such an ellipsoid. However, the problem with this assumption is that the cell does not grow in isolation but is produced as a bud of a mother cell. In Ref. 9, the author acknowledges: "*A yeast cell has only one period of volume growth – while it is being produced as a bud. Thereafter it does not change in size though it continues to produce buds*". This connection does not allow considering the ellipsoid as an adequate geometrical model of the growing bud at the beginning of growth, since the bud and the mother cell are connected. Photos of *S. cerevisiae* shown in Ref. 11 demonstrate this clearly. In fact, both the bud and the mother cell constitute a single system. In Ref. 9 we can read: "*Electron microscope photographs show a cell wall between bud and mother in a mature bud but not in a young bud [1][1]. Also granules can be seen passing from a mother cell into a young bud [2][2]*". So, the actual volume at the beginning is less than it was assumed in Ref. 9. Based on photographs from Ref. 11, we evaluated that the actual volume of a budding cell is about 8-10% lower at the beginning of growth compared to the volume computed in Ref. 9. As the bud grows, this discrepancy quickly decreases, so that except for the two experimental points at the beginning, we can assume that the volume estimation is correct. Fig. 2 shows experimental dependence and the growth curve computed using the growth equation (11).

---

[1] H. D. Agar, H. C. Douglas, *J. Bacteriol.* **70**, 427 (1955).
[2] S. Bayne-Jones, E. F. Adolph, *J. Cellular Comp. Physiol.* **1**, 387 (1932)

The approach we used for finding parameters of the growth equation for *S. cerevisiae* was as follows. First of all, we assume that the rate of growth is not decreasing in the area of the first and the second experimental points. This assumption is well justified. In Ref. 9 the author refers to Ref. 12 as follows: "*They found an increasing rate of volume growth for the first 20 minutes after the appearance of the bud, then linear growth for the next 50 minutes, and finally a declining rate of growth for the last 20 minutes.*" He also refers to Ref. 13 saying this: "*Their results are not unlike those of Bayne-Jones and Adolph, though they emphasize the linear parts of the growth curve*". Of course, there is a possibility that some factor exists that diminishes the growth rate at the beginning, of which we are not aware about. One of the reasons can be that the bud grows in connection with a mother cell, which makes this reproduction somewhat special. However, given the discussed experimental observations and the fact that volume at the beginning in Ref. 9 was not computed correctly, the most likely reason is the aforementioned error in volume calculation that led to volume overestimation at the beginning of growth cycle. Note that this factor does not affect the experimental results above the initial phase of growth, as it was predicted, so that we may rely on experimental data after the first two experimental points. (In fact, if we take into account the third point and correct its position, it will be slightly lowered too, so that the correspondence between the computed growth curve and experiment will be even better, which further reinforces our considerations.)

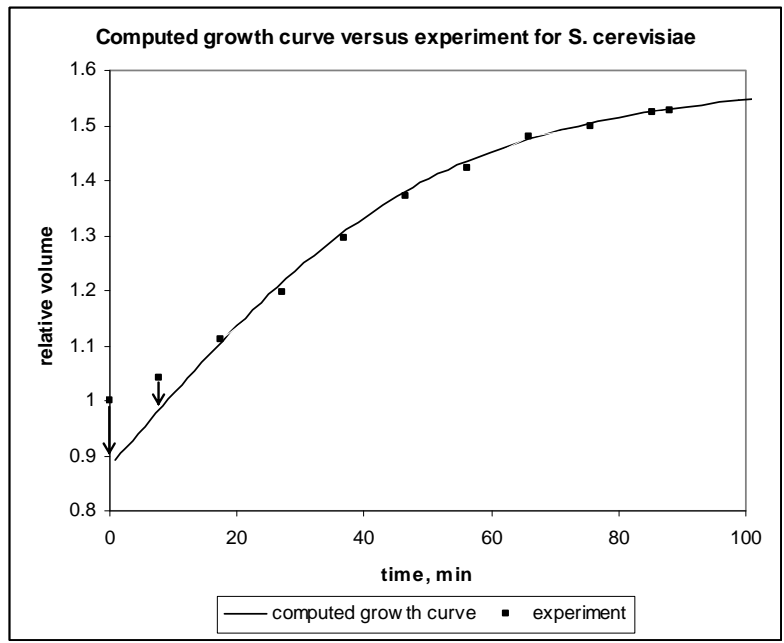

Fig. 2. Computed growth curve versus experimental data from Ref. 9 for *S. cerevisiae*.

We can see that, indeed, the initial part of the growth curve can be approximated by a straight line as the authors of Ref. 13 obtained in experiments. Then, using experimental points 3, 4 and 5, we find the actual beginning volume (relative to experimental data) as 0.91. This is about 9% less then the initial point in Ref. 9. This value corresponds well to our earlier estimation of 8-10% done independently on the basis of geometrical considerations. Since *S. cerevisiae*, as we already found, uses the whole growth cycle, we find the maximum possible volume as follows. First, we normalize the relative volume for the last experimental point, which is equal to 1.527, by the value of 0.91, and then multiply the obtained value 1.678 by 1.02, assuming that the cell's maximum possible volume is 2% greater than the maximum experimentally found volume, the same as for *amoeba*[1]. Then, we use thus obtained value of 1.712 as the maximum possible volume for the growth equation. We finally adjust the computed growth curve vertically relative to experimental points 3, 4 and 5. However, even if don't do that and place the beginning of the computed growth curve at point 0.91, the correspondence of the computed and experimental data will be almost as good. Note that thus computed growth curves have *inflection points*, so that actually the growth curves have some convexity and, as it was found by the authors of works cited above, they are indeed sigmoid curves. However, this convexity is small and the initial part of the growth curve can be well approximated by a line, as some authors of the discussed works did. Overall, except for the first two points, we have a good match between the experimental data and the computed growth curve. Note that the only parameter which we adjusted was a time scaling coefficient.

We also need to find nutrient influxes required for RNA and protein synthesis. If we would model *S. cerevisiae* by a sphere, then we could use influx (10). In case of ellipsoid we do not have analytical solution. So, we should use the general form of growth equation (11), assuming that $K_{min}$ is defined by (6). We do not have data for the values of $C_r$ and $C_p$ in case of *S. cerevisiae*. However, we can evaluate a *range* of these values, based on data for *E. coli* presented in Ref. 14, since *E. coli* also has accelerated rate of RNA synthesis compared to rate of protein synthesis. Because we cover the *whole range* of possible ribosome and protein content, these variations are not meaningful for our purposes. The important thing is to associate the *range* of possible growth rates with appropriate chemical compositions. The summary of data that we use in calculations is presented in Table 2.

Table 2. Chemical composition of *E. coli* cells, [mg/(g dry weight)], from Ref. 14.

| Scenario No. | $\mu$ [1/$hr$] | DNA | RNA | Protein+ tRNA+ | $C_R$ | $C_P$ |
|---|---|---|---|---|---|---|
| 1 | 0.2 | 40 | 35 | 915 | 0.035 | 0.924 |
| 2 | 0.6 | 37 | 90 | 870 | 0.09 | 0.873 |
| 3 | 1.2 | 35 | 135 | 825 | 0.136 | 0.829 |
| 4 | 2.4 | 30 | 250 | 730 | 0.246 | 0.723 |

Formula (6) and appropriate values of $C_R$, $C_P$ for each particular scenario define the nutrient influx to be used in (11). Fig. 2 presents growth curve for scenario 3. Despite the large variation of $C_R$ and $C_P$ coefficients, growth curves for all four scenarios are very close. We did numerical integration using vertical planes that "chop" the horizontally located ellipsoid into vertical slices of equal width along the long axis. In this case, the elementary volume is a difference in volumes of individual slices of two consecutive ellipsoids whose long axes increase by the same step (the same as the layers' width); from the original ellipsoid to the final size. The short axes increase in accordance with the ratios of axes for the current size of ellipsoid.

Fig. 3 shows the change of growth ratio for three forms from Table 1 (for a sphere, an ellipsoid, and an elongating ellipsoid). The growth ratio for an ellipsoid close to sphere is the highest at the beginning.

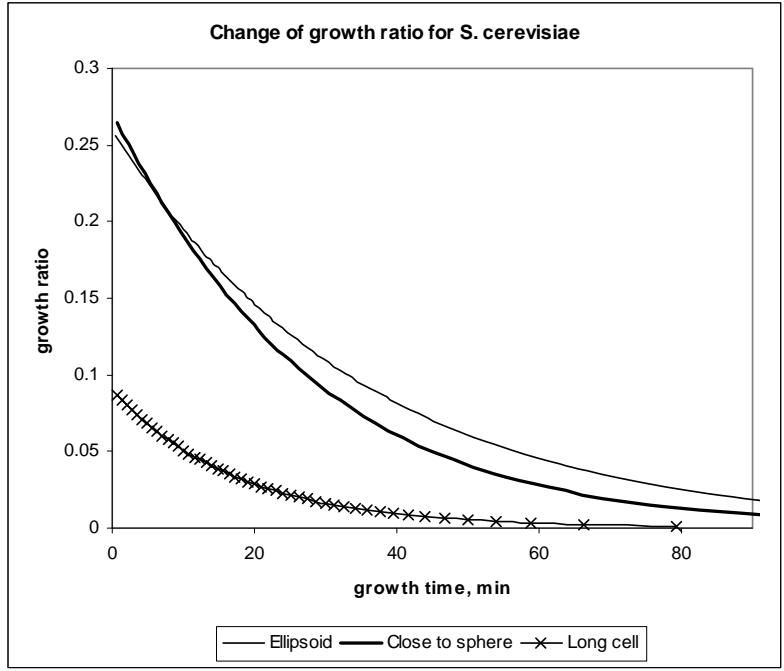

Fig. 3. Change of the growth ratio for *S. cerevisiae* depending on the growth time and geometrical form.

## 5. Finding amount of synthesized biomass for S. cerevisiae

Since the growth ratio defines the amount of nutrients diverted to biomass production, we can find how much biomass *S. cerevisiae* produces during the growth cycle. It is difficult to precisely define which point on the computed growth curve corresponds to the end of growth. We used two approaches. One was based on period of growth. In order to be on the safe side, we computed the amount of produced biomass for a range $6\,\text{min}$, with the whole growth period of about 90 min. The second independent approach was based on the assumption that the division point corresponds to a certain decrease of tangent of slope of growth curve, the same as for the curve that was compared to experimental data (Fig. 2). Both approaches have drawbacks and this is an issue to be studied further. Nonetheless, both of them produced almost identical range of results presented in Table 3.

We considered two ellipsoid forms that double their volume. One was elongating during the growth. An initial ratio of ellipsoid axes was 1.23, and the ending ratio was 1.35. The second ellipsoid increased almost proportionally in three dimensions (accordingly 1.23 at the beginning and 1.24 at the end). The formula that defines amount of synthesized biomass *B* during the growth period is as follows.

$$B = \frac{\int_0^{T_G} G(t)K(t)dt}{\int_0^{T_G} K(t)dt} \qquad (14)$$

where $T_G$ is the total time of growth including division period.

The rationale for (14) is this. Since the value of the growth ratio defines the amount of produced biomass, then the quantity of biomass (as a fraction of the total influx *K*) can be found as an average value of the growth ratio weighted by time and nutrient influx.

Scenario 1, in Table 2, is closer to the case when the dilution rate was 0.1 (1/hr). Scenario 4, also in Table 2, is closer to the situation when the dilution rate was higher, at 0.3 (1/hr). Similar values of produced biomass, computed by methods of metabolic flux analysis (MFA) are taken from Ref. 15. Note that the metabolic flux analysis (MFA) uses many experimental values as inputs, so that the results of MFA modeling are fairly accurate, usually in the range of errors of few tens of percent, depending on the input data and composition of biochemical reactions. In Table 3, we present results for two forms. One is an ellipsoid of revolution with initial and ending ratios of axes accordingly 1.23 and 1.35, that is it elongates during growth. This geometry is close to what can be observed in photos of *S. cerevisiae* in Ref. 11. The second ellipsoid almost preserves its original shape. Both ellipsoids double their initial volumes.

Table 3. Biomass synthesis in percent found using methods of metabolic flux analysis[15] (MFA) and through the growth equation. Figure numbers refer to Ref. 15. DR stands for "dilution rate".

| Method | MFA | axes ratio change is 1.23 – 1.35 | axes ratio change is 1.23 – 1.24 |
|---|---|---|---|
| Basic stoichiometric model (Fig. 3), DR=0.1 (1/hr) | 6.26 | 6.08-6.4 | 6.3-6.7 |
| Basic stoichiometric model (Fig. 4), DR=0.3 (1/hr) | 7.87 | 7.4-7.8 | 7.78-8.1 |
| Catalyzed by ADH III (Fig. 4), DR=0.3 (1/hr) | 8.1 | 7.4-7.8 | 7.78-8.1 |
| Basic stoichiometric model (Fig. 5), DR=0.3 (1/hr) | 7.87 | 7.4-7.8 | 7.78-8.1 |
| Catalyzed by IDP II (Fig. 5), DR=0.3 (1/hr) | 9.08 | 7.4-7.8 | 7.78-8.1 |

## 6. Growth of E. coli and biomass production

We do not have such detailed data for *E. coli* on biomass production. Some qualitative evaluations can be done for the data presented in Ref. 6. In this work, the authors found the biomass yield (amount of produced biomass per certain amount of nutrients): "*If the acetate and glucose consumed are normalized per carbon, the model predicts $Y_{x/s}[g/g(carbon)] = 0.086$*", and compared it versus the cited experimental value of 0.088. This is consistent with the very basic estimate of biomass production for *E. coli*. In particular, the maximum growth ratio is about 0.087 if we assume that *E. coli* grows only in length. If we know the shape of *E. coli* at the beginning and end of growth, we could compute the amount of produced biomass using (14), but we do not have such data. At least we can see that the numbers match in value.

Let us remind that we calculated these values for *evolutionarily* developed cells. In metabolic engineering, the goal is producing substances, so that the biochemical reactions should ideally be set up in such a way that no biomass is produced at all and the cell would not grow to a mature size and divide, while producing the required substance endlessly. So, in industrial strains, one could expect to find smaller values for biomass synthesis and accordingly smaller values of the growth ratio. (Actually, the right causation would be the smaller growth ratio from which the smaller values of biomass synthesis follow.)

## 7. Discussion of results on biomass synthesis for S. cerevisiae

Application of the general growth law to finding synthesized biomass proved its practical usefulness. The computed amounts of biomass accurately correspond to data obtained by methods of metabolic flux analysis. In case of "catalyzed by IDP II" scenario (the last line in Table 3), we have divergence of about 1%. However, this is a special case, for which we do not have reliable input data. The presence of IDP II very likely may affect the shape and / or maximum possible size of *S. cerevisiae* compared to regular growth scenario.

There are two major factors that affect accuracy of this method. One is the shape of a growing organism, and the other one, which is largely a computational issue, is determining the end of growth for the computed growth curve. We made reasonable assumptions about the shape of *S. cerevisiae* and estimated the range of possible variations in biomass production due to form change. It was found that such variations produce close values, so that with regard to a form variation the method is stable. We proposed two approaches for finding an end of growth period that worked well in our case. However, since cells experience significant biochemical changes in the division phase, this study should continue and maybe more interesting facts and relationships will be found. For instance, it will be interesting to study specific properties of nutrient influx right before and during the division phase.

Combining the proposed method and the present stoichiometric approaches will definitely improve methods of metabolic flux analysis, in terms of simplification, stability and accuracy of solutions. In metabolic flux analysis, the amount of produced biomass is unknown. Most often, the solution of system of stoichiometric equation is optimized for production of maximum amount of biomass. Once the amount of synthesized biomass is known, which the proposed method allows to do independently, the problem becomes significantly simpler in mathematical and computational terms. We can see from comparison of data in Table 3 that results obtained on the basis of the growth equation match computations done by methods of metabolic flux analysis fairly well, given the uncertainty of the actual form of *S. cerevisiae* and its change during growth, as well as presence of errors introduced by metabolic flux analysis methods. Besides, MFA computes the amount of produced biomass for a certain composition of biochemical reactions, while we compute the amount of biomass produced during the *whole* growth cycle. (Of course, the method can be used for finding amount of synthesized biomass at any given moment of growth since it is defined by the value of the growth ratio.)

An additional indirect proof of validity of our computations of biomass production is that for a *lower* concentration of nutrients (scenario 1 from Table 2) we accordingly obtained *lower* value of synthesized biomass compared to scenario 4 with more nutrients, which is in good agreement with results obtained by MFA approach used in Ref. 15.

Overall, the estimation of biomass production based on the growth equation is consistent with available experimental data and results produced by stoichiometric models used in metabolic flux analysis. However, the proposed approach on the basis of the growth equation is much simpler and, in fact, more efficient, since it allows easily differentiating and finding biomass production for any stage of growth, both for individual organisms and their colonies in bioreactors. Stoichiometric methods, in this regard, are much more complicated methods, especially when it comes to description of dynamical growth processes. On the other hand, stoichiometric methods can benefit from the proposed method very much. Solution of systems of stoichiometric equations is based on solving an optimization problem, when the solution is optimized for a maximum of produced biomass. Once we know how much biomass is produced, which can be done using the proposed method, we significantly simplify the problem in mathematical and computational terms. Besides, usage of the growth equation allows introducing additional constraints that potentially can improve the accuracy and stability of stoichiometric solution.

## 8. Application of results in biotechnology and bioengineering

### 8.1. Balanced growth

Based on the above results, one evident application of the general growth mechanism in biotechnology and bioengineering is evaluation of biomass production and incorporation of this information into the methods used by these disciplines, which can be done in a mathematical form via additional constraints and equations. It was shown in many publications, such as Ref. 14, that biomass production is one of the most definitive factors in the growth and replication processes. For instance, systems of stoichiometric equations whose optimization is based on maximization of biomass production produce substantially better results compared to other stoichiometric criteria, which is a proof of the adequacy of this approach. Note that the new information that can be derived from the growth equation can be incorporated both into underdetermined, overdetermined and well defined systems of appropriate equations. For instance, when the growth is balanced, we can assume that the concentrations of intracellular metabolites are constant, so that

$$S \cdot v = b \qquad (15)$$

where $S$ is a stoichiometric matrix (with a dimension $n \times m$); $v$ is a vector of reaction rates (or fluxes); $b$ is the vector of consumption and secretion rates of metabolites and biosynthetic requirements for cellular macromolecules[6, 16]. The number of components is substantially less than the number of reactions. Then, in order to solve such an underdetermined system of stoichiometric equations, one can use a certain objective function, applying it toward the minimization or maximization of

$$Z = \sum_i c_i v_i \qquad (16)$$

Here, $c_i$ are the weights and $v_i$ are the elements of the flux vector, and $Z$ is the amount of produced biomass. In fact, this solution maximizes the synthesis rate of each precursor for biomass composition, such as individual amino acids, nucleotides, etc. In other words, flux balance analysis optimizes the set of fluxes such that the flux through a particular cellular reaction is maximized (or minimized)[16]. Common choices of cellular objective functions in models of metabolic networks include biomass production, energy and byproduct production[16]. The maximization of the growth rate, or biomass production, is a very common and the most adequate approach for such problems. This concept is in full compliance with the general growth law, which we discussed above.

We can also use the earlier introduced methods of finding the amount of produced biomass on the basis of the growth equation in order to solve (15). In this case, instead of solving the optimization problem (16), which assumes that the amount of synthesized biomass is unknown, we could solve a mathematically less complicated problem, because the value of the synthesized biomass is already known. In other words, instead of solving optimization problem (16) we should find the solution of (15) (unknown values of $c_i$) using the following condition.

$$\sum_i c_i v_i = Z_0 \qquad (17)$$

where $Z_0$ is now known; it is found on the basis of growth equation using methods proposed in this article.

Such an approach would have a positive influence on the mathematical, computational aspects of the problem and accuracy and stability of solution. Indeed, instead of an optimization problem, which requires lots of computational resources and advanced approaches, we have a mathematically better defined problem, because we need to find the solution that satisfies an already known amount of produced biomass. Since we added new information, our solution *potentially* could be more accurate if we use the right mathematical methods.

How successfully this potential can be realized depends on the efficiency of mathematical and computational methods one uses to solve the problem. Generally, solutions of systems of linear equations, such as (15), have an improved accuracy and stability when one adds new independent constraints and equations, so that if one uses more or less optimal methods for solving (17), the chances are very good that the solution will substantially improve.

### 8.2. Dynamic growth

In the Ref. 17, the authors rightly pay attention to the fact that growth processes are dynamic in nature. For that reason, the constraints applied to the system of stoichiometric equations that describe the growing organism change in time, and this should be taken into account. In this regard, the proposed approach finding biomass production inherently takes into account the growth dynamics through the changing growth ratio and other parameters of the growth equation. For dynamic flux analysis, the usage of the general growth mechanism brings enormous advantages from all perspectives, including mathematical, computational and solutions' accuracy and stability. For instance, in case of using the dynamic optimization approach, we have the following advantages. On the mathematical side, instead of solving complicated non-linear programming (NLP) problem, we would have to solve a system of linear differential equations with well defined constraints and boundary and initial conditions, which is much easier than solving the NLP problem. Since the general growth mechanism introduces additional information in the form of additional equations, the accuracy and stability of solution generally has to significantly improve as well. In case of using static optimization approach, introduced in Ref. 17, we also significantly benefit from the usage of the general growth equation and the growth ratio. In this case, besides the advantages already listed for the dynamic optimization approach, we can add additional conditions on the boundaries of time intervals for the amount of produced biomass, since through the general growth mechanism we can find the amount of produced biomass without solving the optimization problem. This is a big advantage from all perspectives.

**8.3. Other flux analysis and bioengineering applications**

Similar to the examples above, the amount of produced biomass can be used as additional information for many other methods in metabolic flux analysis and other areas of biotechnology in order to substantially improve the accuracy and stability of solutions.

Another promising area of application of the general growth mechanism in biotechnology is this. By and large, there is much similarity between biochemical mechanisms across different organisms, such as, for example, the backbone reactions for the ATP. Since the amount of produced biomass is the major parameter to which the composition of biochemical reactions is tied to, once we know the amount of produced biomass, in theory, we can make the backward transition to the composition of biochemical reactions. The approach can be as such. First, we consider the backbone biochemical reactions typical for the considered class of organisms, or maybe even a wider group. Then, the second echelon of biochemical reactions can be considered in order to adjust the overall biomass production to the value defined by the growth equation. Overall, the usage of the general growth mechanism and the growth equation could significantly improve the knowledge of biochemistry of living organisms and uncover many biochemical fine tuning mechanisms, which would allow significantly improved control over the biochemical reactions.

Another important aspect is that the general growth mechanism is like a common denominator that affects all other growth and replication mechanisms and their evolutionary development. Knowledge of evolutionary development is a very potent cognition instrument that gives a developmental perspective to the specific biochemical and biophysical properties of certain organisms. In fact, the general growth law governs the *growth* and *evolution* of all living species, so that understanding the fundamental nature of the growth phenomena through the general growth mechanism opens lots of new opportunities for enhancing control of biochemical machinery of living organisms and development of new bioengineering methods. Understanding the general growth mechanism is like acquiring a bird's eye view of the growth phenomena in general.

**8.4. Secretion of industrial substances**

This is an important subsection that generalizes the application of the general growth mechanism not only to evolutionarily developed organisms, but to all organisms that are used for secretion of certain substances in industrial and other applications. The readers could get an impression from the presented material that the general growth mechanism exclusively describes naturally developed organisms, whose evolution required fast growth and consequently the fastest possible biomass production. In fact, the general growth mechanism is universally applicable to *all* living organisms, regardless of their origin, biological modifications and the phase of growth or existence. The reason for such possible misunderstanding could be the use of the objective function implementing the maximum growth rate in our earlier examples. In fact, the situation is much more interesting and in full compliance with the general growth law.

Suppose that some microorganism is used for secretion of a certain substance. The primary goal of a bioengineer in this case would be the maximum production of the required substance and certainly not the growth of the organism. In fact, the best outcome would be if this organism, once it reaches a certain size, continues secretion of the required substance endlessly. This is an extreme situation. However, it is interesting to see how the general growth law is applied in this case, so that we could explore the range of its applicability. The growth equation is applicable to this situation as well. The current size of the grown organism can be used for the evaluation of the maximum possible size that should be substituted into the growth equation. The value of the produced biomass will accordingly be zero, while the overall composition of biochemical reactions would be arranged in such a way that, besides the other synthesized substances, the output of the secreted substance can be found by conventional methods, let us say, on the basis of stoichiometric equations. As we could see above, using the growth equation, we can define a new objective function in the problem of finding the composition of biochemical reactions for particular organisms. We have introduced additional equations that can be interpreted as additional constraints. Note that we solve no optimization problem, which today is the foundation of stoichiometric methods. This is an important consideration that *fundamentally* changes the approach toward the study of the biochemical machinery of living organisms.

## 9. S. cerevisiae as an example of evolutional development. How organisms provide fast growth

*S. cerevisiae* can serve as example of how organisms adapted to evolutionary requirement of providing fast growth. As we found, this organism uses the whole growth cycle predefined by the growth equation, like *amoebae*. However, the whole growth cycle is not optimal from the perspective of fastest growth[1]. And nonetheless, *S. cerevisiae* grows fast. In fact, there is no contradiction in such behavior. Recall that *S. cerevisiae* has a geometrical form which is close to spherical. As it was shown in Refs. 7, 18, 19, from the geometrical perspective, a spherical form provides the fastest growth among all forms, provided all of them have the same specific influx. The growth ratio for a sphere at the beginning is about two times larger than the growth ratio of a short cylindrical cell with initial length one and a half of diameter. This means that the amount of nutrients diverted in *S. cerevisiae* to biomass production is accordingly greater at the beginning of the growth cycle by about two times compared to a cylinder like organism. Although the growth ratio of a spherical cell decreases faster than in case of elongated forms, the intensive production of biomass at the initial phases of growth gives such cells an advantage in terms of growth rate. This is one factor that increases the growth rate of *S. cerevisiae*, despite the fact that it uses the full growth cycle.

Another factor is that the rate of RNA synthesis is twice as large as the rate of protein synthesis. So, *S. cerevisiae*'s production capacity of protein grows much faster, than, for instance, in the case of *amoeba*. This, accordingly, also allows *S. cerevisiae* to grow fast.

There is also a possibility of change of the geometrical form of *S. cerevisiae* in such a way that it might facilitate faster growth. Although this effect may not necessarily apply to *S. cerevisiae*, we should not rule out the possibility given the variability of *S. cerevisiae* forms[11, 20]. Such variability means that this organism has mechanisms for changing the geometrical form, which can be used for accelerating the growth rate.

Thus, by combining different advantageous growth mechanisms, *S. cerevisiae* is able to grow fast, in spite of using the whole growth cycle.

From the developmental perspective, we would also like to mention that *S. cerevisiae* is one of the oldest organisms. Its growth and replication mechanisms are based on the most fundamental principles, which are a direct consequence of the general growth law;

we can call them the "first line" growth mechanisms. In case of *S. cerevisiae*, these are the round shape that secures a high growth ratio at the beginning, the full growth cycle predefined by the growth equation, and the simpler division mechanisms founded entirely on change of the growth ratio[1]. More sophisticated growth scenarios, such as a division at inflection point, which is the case of *S. pombe, E. coli*, could be developed *only on top* of this primordial growth and replication machinery, as the "second line" of evolutionary growth mechanisms. For instance, *S. pombe* acquired more elaborated division mechanisms, adjusting the beginning of its division phase to the inflection point of the growth curve and adding more biochemical mechanisms that support this earlier division, such as the change of spatial gradient of Pom1 cyclin. Similarly, the double rate of RNA synthesis compared to protein synthesis, which *S. cerevisiae* acquired during evolution, is an enhancement on top of more ancient organisms that synthesized their components at the same rate.

It is known that *S. pombe* and *S. cerevisiae* branched about 300-600 million years ago, but which one is closer to their predecessor? Application of the general growth law, based on the above considerations, answers this question, that it is *S. cerevisiae*, or some organisms very similar to it, which were very likely *direct* ancestors of *S. pombe*.

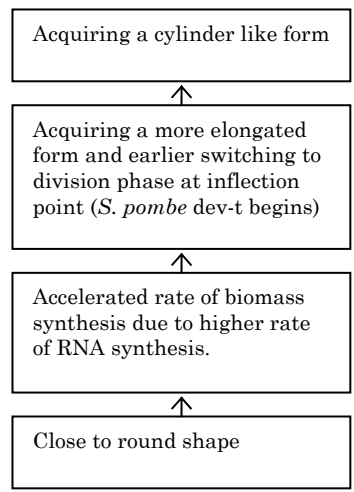

Fig. 4. Evolutionary development of features of *S. cerevisiae* and *S. pombe* which provide their fast growth.

A summary and sequence of evolutionary development of certain features and mechanisms in *S. cerevisiae* and later *S. pombe* which provide their fast growth, is presented in the diagram in Fig.4. We can see how more sophisticated accelerated growth mechanisms are built on the basis of previously developed more primitive mechanisms.

**Acknowledgements**

The author thanks Dr. Piotr Pawlowski and Dr. Peter Fantes for the ongoing support of this study, and Alexander Shestopaloff for numerous fruitful discussions and editing efforts.